\documentclass[12pt,preprint]{emulateapj}

\usepackage{natbib}
\def\sun{\hbox{$\odot$}}

\def\HST{{\it HST}}
\def\kms{{\>\rm km\>s^{-1}}}

\shorttitle{Evolution of V605 Aql}
\shortauthors{Clayton et al.}

\begin{document}

\title{Evolution of the 1919 Ejecta of V605~Aquilae\altaffilmark{*,$\dag$}}

\author{Geoffrey C. Clayton\altaffilmark{1},
Howard E. Bond\altaffilmark{2,3,4}, 
 Lindsey A. Long\altaffilmark{5}, 
Paul I. Meyer\altaffilmark{5}, 
Ben E.~K. Sugerman\altaffilmark{5},
Edward Montiel\altaffilmark{1},
William B. Sparks\altaffilmark{2},
M.~G. Meakes\altaffilmark{2},
O.~Chesneau\altaffilmark{6}, and O. De~Marco\altaffilmark{7}  
}

\altaffiltext{1}{Dept.\ of Physics \& Astronomy, Louisiana State University,
Baton Rouge, LA 70803; gclayton@fenway.phys.lsu.edu, emonti2@lsu.edu}

\altaffiltext{2}{Space Telescope Science Institute;  3700 San Martin Drive,
Baltimore, MD 21218; bond, sparks@stsci.edu,mgmeakes@gmail.com}

\altaffiltext{3}{Department of Astronomy \& Astrophysics, Pennsylvania State University,
University Park, PA 16802}

\altaffiltext{4}{Current address: 9615 Labrador Ln., Cockeysville, MD 21030}

\altaffiltext{5}{Dept.\ of Physics and Astronomy, Goucher College, 1021 Dulaney
Valley Rd., Baltimore, MD 21204; ben.sugerman@goucher.edu}

\altaffiltext{6}{Observatoire de la C\^{o}te d'Azur-CNRS-UMR 6203, Dept Gemini,
Avenue Copernic, F-06130 Grasse, France; olivier.chesneau@ob-azur.fr}

\altaffiltext{7}{Dept.\ of Physics, Macquarie University, Sydney, NSW 2109,
Australia; \hbox{orsola}@science.mq.edu.au}

\altaffiltext{8}{European Southern Observatory, Karl-Schwarzschild-Strasse 2
D-85748 Garching bei M{\"u}nchen, Germany; fkerber@eso.org}

\altaffiltext{*}
{Based on observations made with the NASA/ESA {\it Hubble Space Telescope},
obtained by the Space Telescope Science Institute, and from the data archive at
STScI\null. STScI is operated by the Association of Universities for Research in
Astronomy, Inc., under NASA contract NAS5-26555.}

\altaffiltext{$\dag$}{We dedicate this paper to the memories of William P. Bidelman (1918--2011) and Martha L. Hazen (1931--2007)---teachers, colleagues, friends, lovers of the stars.}

\begin{abstract}
New imaging of V605 Aql, was obtained in 2009 with {\it HST\/}/WFPC2, which
had a nova-like outburst in 1919, and is located at the center of the
planetary nebula (PN), Abell 58. This event has long been ascribed to a final helium shell flash, but it has been suggested recently that it may instead have been an ONe nova.  The new images provide an 18 year baseline for the expansion of the ejecta from the 1919 event. In
addition, the central star has been directly detected in the visible
for the first time since 1923, when it faded from sight due to
obscuration by dust. The expansion of the
ejecta has a velocity of $\sim$200 km s$^{-1}$, and an angular
expansion rate of $\sim$10 mas~yr$^{-1}$, consistent with a 1919 ejection. This implies a geometric distance of 4.6 kpc for V605 Aql,
consistent with previous estimates. The gas mass in the
central knot of ejecta was previously estimated to be
$5\times 10^{-5}\ M_{\sun}$. It is estimated that warm dust
associated with this gas has a mass of $\sim$$10^{-5}\
M_{\sun}$. There is also evidence for a significant amount,
$10^{-3}\ M_{\sun}$, of cold (75 K) dust, which may be
associated with its PN. The knot ejected in 1919 is asymmetrical and
is approximately aligned with the asymmetry of the surrounding
PN. 
Polarimetric imaging was obtained to investigate whether the 2001
spectrum of V605 Aql was obtained primarily in scattered light from
dust in the central knot, but the signal-to-noise in the data was
insufficient to measure the level of polarization.

\end{abstract}

\keywords{stars: individual (V605 Aql)}

\clearpage

\section{Introduction}

In the second decade of the 20th century, the variable star
V605~Aquilae underwent a remarkable outburst
\citep{1997AJ....114.2679C,2006ApJ...646L..69C}.  Its eruption was
discovered by \citet{1920AN....211..119W} on plates taken at
Heidelberg, with the star reaching a maximum photographic magnitude of
10.2 in 1919 August. Subsequent investigation of Harvard plates by Ida
Woods showed that V605~Aql had slowly risen to its maximum over the
previous 2~years \citep{1921BHarO.753Q...2B}. The historical visual
and photographic photometry has been assembled and discussed by
\citet[][see also
  \citet{1996PASP..108.1112H}]{2002Ap&SS.279....5D}. Although
initially considered to be a very slow nova---it was designated Nova
Aquilae No.~4---V605~Aql had a light curve unlike that of any known
nova. After fading in 1920, it re-brightened in 1921, faded again, and
then brightened yet again in 1923, before finally fading from
sight. \citet{1981MitVS...9...13F} examined over 400 plates obtained
between 1928 and 1979, and found that V605~Aql was always below
photographic magnitude, 16--17.5, during this interval. Spectroscopy
obtained in 1921 September, showed the object to be a cool carbon
star, with a spectral type of R0, in stark contrast to the
high-excitation emission-line spectra of classical novae at late
outburst stages \citep{1921PASP...33..314L}.

In spite of these unique features, V605 Aql was almost entirely
ignored by astronomers over the ensuing five decades. Modern interest
was rekindled by \citet{1971ApJ...165L...7B}, who reviewed the scanty
data and made the novel suggestion that V605~Aql had been an
extragalactic supernova (SN)\null. His speculation was based on
Lundmark's R0 classification along with the fact that the first SN
whose spectrum was photographed, Z~Centauri (SN~1895B), had a spectrum
that had been considered to resemble that of an R-type star by
\citet{1916AnHar..76...19C}. However, \citet{1973BAAS....5..442B}
subsequently was able to examine Lundmark's original 1921
spectrograms, and verified that V605~Aql in outburst indeed was a
carbon star---in fact, a {\it hydrogen-deficient\/} carbon (HdC) star,
closely similar to the well-known non-variable HdC star HD~182040, as
well as other cool R Coronae Borealis (RCB) stars
\citep{2012JAVSO..40..539C}. The 1921 spectrum was re-examined and
presented in \citet{1997AJ....114.2679C}.

Following up on \citet{1971ApJ...165L...7B},
\citet{1971PASP...83..819V} obtained deep photographs of V605~Aql with
the 5-m Hale telescope. These images showed it to be the central star
of a large, very faint, old planetary nebula (PN), which had already
been cataloged as object~58 in the \citet{1966ApJ...144..259A} list of
faint PNe discovered in the course of the Palomar Observatory Sky
Survey (POSS)\null.  Abell~58 (PN~G037.5$-$05.1) has an elliptical
shape, with dimensions of $44''\times36''$. See Figure~1.

\begin{figure}
\begin{center}
\includegraphics[width=3.0in,angle=0]{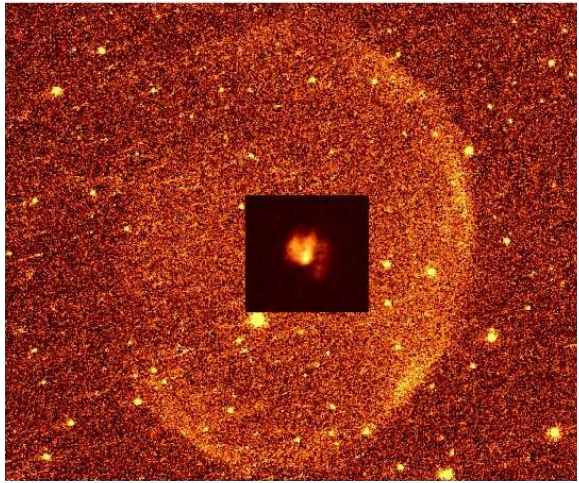}
\end{center}
\figcaption{Combined F658N  ([N II]) {\it HST\/WFPC2} image from the Hubble Legacy Archive with an inset showing a magnified image of the central knot. The relative orientations of the inner knot and the outer PN are similar. North is up and East is to the left and the size of the image is approximately 50\arcsec x 50\arcsec. The inset is  5\arcsec x 5\arcsec.}
\end{figure}

Based on these findings---a cool, hydrogen-deficient star erupting in
the center of an old PN---many investigators believe V605~Aql to be a
star that underwent a final helium-shell flash (FF), an event first
described theoretically by \citet{1983ApJ...264..605I}.  In the FF
scenario, a hot PN nucleus (PNN) re-ignites He shell burning, retraces
its evolution to become temporarily a ``born-again'' red giant whose
remaining surface hydrogen has been ingested and burned, and then
evolves back to its original high surface temperature.  Recent
theoretical models indicate that the transition from a hot PNN to a
cool giant, and then back again, can occur in as little as 5--10~yr.
Such extremely rapid evolution results from a very late thermal pulse,
in which the star experiences a FF only when it has already reached
the top of the white-dwarf cooling track with a very thin envelope
remaining
\citep{2001ApJ...554L..71H,2003ApJ...583..913L,2005Sci...308..231H}.

V605~Aql in 1921 had a spectrum resembling that of a cool RCB star
with $T_{\rm eff}\simeq5000$~K \citep{1997AJ....114.2679C}. However,
there is evidence that it had reheated no later than the mid-1970s,
and likely as early as the early 1950s, if not even earlier. At
present, a dusty knot of extremely hydrogen-deficient nebulosity,
unresolved or marginally resolved in ground-based optical images, lies
at the site of V605~Aql at the center of the large, faint PN, Abell~58
\citep{1997AJ....114.2679C,2006ApJ...646L..69C}. The spectrum of the
compact knot is extraordinary. It shows strong emission lines of
[\ion{O}{3}], but no sign of the Balmer lines
\citep{1985MitAG..63..181S}. The surrounding PN, however, has a fairly
normal hydrogen-rich composition
\citep[e.g.,][]{1971ApJ...170..547F,2008MNRAS.383.1639W}.  Optical
spectra of the knot, obtained in 2001, confirm the presence of broad,
high-excitation Wolf-Rayet stellar features, including strong
\ion{C}{4} emission, and clearly demonstrate that the star is again
very hot, $\sim$95,000 K \citep{2006ApJ...646L..69C}. Since the knot
is visible in blue POSS images obtained in 1951, V605~Aql was most
likely already producing ionizing radiation and strong [\ion{O}{3}]
emission by that time. A spectrogram obtained in the mid-1970s by
P.~Osmer with the Cerro Tololo 4-m telescope at the request of
H.E.B. showed the nebular emission lines, but unfortunately the
unpublished data have now been lost. The compact knot is probably
material ejected in the 1919 outburst, which is so dusty that it hides
the star from direct view; possibly we see only starlight that emerges
after scattering off the dust \citep{2006ApJ...646L..69C}.

The behavior of V605~Aql is very similar to that seen more recently in Sakurai's
Object (V4334~Sgr), which also lies within a very faint PN\null. After a rapid
brightening in 1996, followed by several fluctuations, Sakurai's Object went
into a deep decline in 1999, just as V605~Aql did in 1923 
\citep{1999A&A...350L..27K,2002Ap&SS.279....5D}.
Thick  dust formed around both stars, causing their precipitous
optical declines. The spectrum of Sakurai's Object in 1997 was very similar to
the 1921 spectrum of V605~Aql \citep{2001Ap&SS.275...91K}. Also, Sakurai's Object shows
evidence for new circumstellar ionized gas only 10 years after its outburst
\citep{2002ApJ...581L..39K,2005Sci...308..231H}, suggesting that it is already
becoming hot again. Thus, the current behavior of V605~Aql is a template for the
future behavior of Sakurai's Object. Another FF object, also surrounded by an old
PN, FG~Sge, is evolving much more slowly than the near-twins V605~Aql and
Sakurai's Object 
\citep{2002Ap&SS.279....5D,2004A&A...426..145L}.

Recently, however, the FF scenario for V605 Aql has been questioned by 
\citet{2008MNRAS.383.1639W} and \citet{2011MNRAS.410.1870L}. They carried out a spectroscopic analysis of
the V605~Aql central knot, showing a C/O ratio below unity and an remarkably high abundance of Ne,
properties that are in stark conflict with theoretical models of FF events. These
findings suggest a more complicated evolutionary history, involving a ONe
white dwarf involved either in a binary merger or in an unusual classical-nova
outburst.

In this paper, we present new polarimetric high-resolution imaging of V605~Aql with the {\it
Hubble Space Telescope\/} (\HST\/).
The polarimetry was done to investigate whether
the recent spectrum of V605 Aql is seen only in scattered light, in which case the knot surrounding the star should be highly polarized. These new images also provide an 18 year baseline to measure the expansion and evolution of the central knot. 
We examine the morphology
and angular expansion of the compact nebula, estimate its distance and reddening, while
discussing the implications for the nature of V605 Aql.


\begin{deluxetable*}{lllccl}
\label{t:absphot}
\tablewidth{0 pt}
\tablecaption{{\it HST\/} Imaging of V605~Aquilae used in this study}
\tablehead{
\colhead{Date [UT]} &
\colhead{Instrument} &
\colhead{Filter} &
\colhead{Image Scale} &
\colhead{Exposure} &
\colhead{Program ID/PI} \\
\colhead{} &
\colhead{} &
\colhead{} &
\colhead{[$''$/pix]} &
\colhead{[s]} &
\colhead{} 
}
\startdata
1991 Aug 6   & FOC	 & [\ion{O}{3}] F501N & 0.014 & 796  & 2570/Bond \\
\noalign{\vskip0.125in}
2001 May 27  & WFPC2/WF3 & [\ion{O}{3}] F502N & 0.100 & 1050 & 9092/Hinkle \\
2001 May 27  & WFPC2/WF3 & [\ion{N}{2}] F658N & 0.100 & 460  & 9092/Hinkle \\
\noalign{\vskip0.125in}
2009 Mar 9   & WFPC2/PC  & [\ion{O}{3}] F502N & 0.046 & 800  & 11985/Clayton \\
2009 Mar 9   & WFPC2/PC  & [\ion{N}{2}] F658N & 0.046 & 1200 & 11985/Clayton \\
2009 Apr 3-5 & WFPC2/WF2 & \ion{C}{4}	F547M & 0.100 & 12,600 & 11985/Clayton \\
             &           & \quad + polarizers \\        
\enddata
\end{deluxetable*}

\section{Observations and Data Reduction}
\subsection{\HST\/}

V605~Aql has been imaged at three epochs by \HST, as summarized in
Table~1.  Observations were made at epoch 1991.60 (PI: H.E.B.) with
the aberrated Faint Object Camera (FOC) at $f/96$ in two broad-band
ultraviolet filters, an optical blue continuum filter (F437M), and a
narrow-band filter isolating [\ion{O}{3}] 5007~\AA\null. The compact
nebula was detected only in the [\ion{O}{3}] exposure, listed in
Table~1. Brief discussions of the 1991 images were given by
\citet{1992ESOC...44..139B,1993IAUS..155..499B} and
\citet{2002Ap&SS.279...31B}, and the images were described in more
detail by \citet{1997AJ....114.2679C}. Second-epoch observations were
obtained at 2001.40 (PI: K.~Hinkle) using the Wide-Field 3 chip (WF3)
of the Wide Field Planetary Camera~2 (WFPC2) and filters isolating
[\ion{O}{3}] 5007~\AA\ and [\ion{N}{2}] 6583~\AA\null. Details of
these observations are given by \citet{2008A&A...479..817H}, who also
presented high-resolution ground-based near-IR images.

The new third-epoch imaging, reported here, were obtained at epoch
2009.21 (PI: G.C.C.). Narrow-band WFPC2 frames in the nebular
[\ion{O}{3}] 5007~\AA\ and [\ion{N}{2}] 6583~\AA\ emission lines were
obtained with the PC1 chip, making these the highest-resolution
optical images available for V605~Aql.

We also used WFPC2 to obtain polarimetric images at epoch 2009.3 in the
medium-band F547M filter, which in the case of V605~Aql transmits the broad
Wolf-Rayet \ion{C}{4} 5806~\AA\ band emitted in the wind of the central star,
and mostly rejects emission from the surrounding nebula. Thus the F547M signal
is expected to be dominated by direct light from the central star, plus any
component of starlight scattered by the surrounding dust before emerging from
the compact nebula, which would be expected to be linearly polarized.

For the polarimetric imaging, we used ``Strategy 2c,'' as outlined by \citet{1995wfpc.rept....1B}. The technique is to place the target in a corner of the WF2 chip
near the amplifiers (for less degradation by charge-transfer inefficiency in the
aging detector) and make exposures with four polarizer rotations, denoted POLQ,
POLQP15, POLQN33, and POLQN18. These yield respective electric-vector
orientations of $0^\circ$, $15^\circ$, $102^\circ$, and $117^\circ$, relative to
the telescope orientation angle. The exposures were made during three two-orbit
visits, accumulating totals of 3100 or 3200~s in each of the four polarizer
settings.

Pipeline-calibrated \HST\/ images were retrieved from the archive and
processed as follows: (1)~The 1991 FOC [\ion{O}{3}] image, obtained
with the aberrated telescope, was deconvolved using 80 iterations of
the Lucy-Richardson procedure in the IRAF\slash STSDAS\footnote{IRAF
  is distributed by the National Optical Astronomy Observatories,
  which are operated by the Association of Universities for Research
  in Astronomy, Inc., under cooperative agreement with the National
  Science Foundation. The Space Telescope Science Data Analysis System
  (STSDAS) is distributed by the Space Telescope Science Institute.}
package. Pictorial representations of the raw and deconvolved images
have been presented by \citet{2002Ap&SS.279...31B}.  (2)~The WFPC2
images from 2001 and 2009 were combined using {\tt multidrizzle}
(within the STSDAS package) to the native pixel scales of the
wide-field (2001; $0\farcs1\,\rm pix^{-1}$) and planetary (2009;
$0\farcs045\,\rm pix^{-1}$) chips. In order to compare the 2001 and
2009 epochs directly, we also drizzle-combined the latter to the 2001
WF2 resolution of $0\farcs1\,\rm pix^{-1}$. Field stars, common to all
of the frames, were then used to compute the geometric transforms
necessary to register all of the images to better than 0.1~pixel rms.

For the polarimetric analysis of the surrounding compact dust nebula,
we manually registered the four individual images using
cross-correlation, median combination, and sky subtraction. The final
stacks for each polarization angle were geometrically registered to a
common reference frame in the same manner as for the narrow-band
images. The Stokes parameters $U$, $Q$, and $I$ were then computed
from linear combinations of three polarization-angle images at a time,
as outlined using the online WFPC2 Polarization Calibration
Tool\footnote{http://www.stsci.edu/hst/wfpc2/software/wfpc2\_pol\_calib.html}.
To increase the signal-to-noise ratio, at the expense of some
resolution, we block-averaged the images by a factor of two, computed
Stokes-parameter images for all four permutations of the polarization
images, and averaged the resulting data to produce final Stokes
images.  From these, we calculated the degree of linear polarization
and the position angle of the electric vector at each pixel.

To verify our procedure, we also determined $P$ for several
unsaturated field stars in the neighborhood of the nebula. They indeed
have negligible values of $P$, as expected since interstellar
polarization is almost always quite small.  Propagating the Poisson
and read-noise errors associated with each individual frame, and the
variance of the individual Stokes images, we find these detections of
polarization significant at roughly a 2$\sigma$ level. While there is
some indication that the nebula may have significant levels of linear
polarization, the signal-to-noise ratio of these data is too low for
further analysis.

\begin{figure}
\centering
\includegraphics[height=\linewidth,angle=-90]{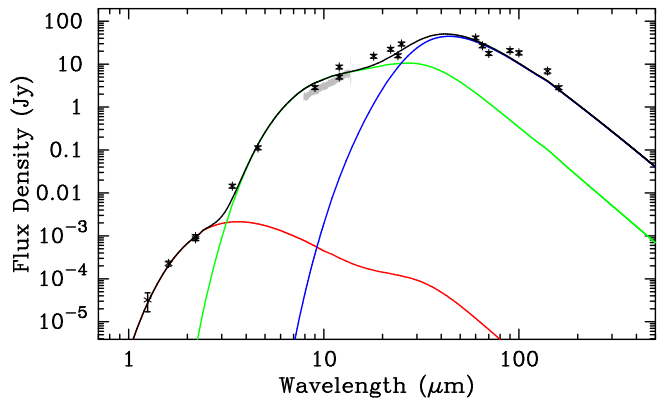}
\caption{The SED of V605 Aql is plotted including $JHK$ photometry
  \citep{2008A&A...479..817H}, and photometry from {\em IRAS}, {\em
    AKARI}, {\em WISE}, and {\it Spitzer}/MIPS. The data are listed in
  Table 2. The grey line is a spectrum obtained with MIDI/VLTI.  A
  three component fit has been done to the points using amorphous
  carbon dust and three temperatures, 810 K (red), 235 K (green), and
  75 K (blue). The sum of three components is shown as a black line. \\
}
\end{figure}

\subsection{Infrared Photometry}

The spectral energy distribution (SED) of V605 Aql shows that there is
significant dust around the star emitting in the mid- to far-IR. The
SED is plotted in Figure~2. The data, summarized in Table 2, consist
of JHK photometry \citep{2008A&A...479..817H}, as well as photometry
from the {\it Infrared Astronomy Satellite} ({\em IRAS})
\citep{Walker:1985rr, 1986MNRAS.222..357K}; the {\em AKARI} Infrared
Camera (IRC) and Far-Infrared Surveyor (FIS)
\citep{2007PASJ...59S.369M,2010A&A...514A...1I}; the {\em Wide-field
  Infrared Survey Explorer} ({\em WISE}) \citep{2010AJ....140.1868W};
and the {\it Spitzer} Multiband Imaging Photometer for Spitzer (MIPS)
\citep{2004ApJS..154...25R}. MIPS scans at 24, 70, and 160 \micron~of
V605 Aql were downloaded from the archive and photometry was done
using PSF-fitting in Starfinder. The other IR Photometry of V605 Aql
was collected from the literature and from Vizier.

\begin{deluxetable}{ll}
\tabletypesize{\scriptsize}
\tablewidth{0pt} 
\tablecaption{V605 Aql Photometry}
\tablenum{2}
\tablehead{\colhead{Band}&
           \colhead{Flux (Jy)}}
\startdata

$J^a$ &  3.21e-05 $\pm$ 1.5e-05\\
$H^a$&2.27e-04  $\pm$ 3.0e-05\\
$K^a$& 8.56e-04   $\pm$  8.0e-05\\
{\em IRAS}/12&	4.99e+00  $\pm$ 2.0e-01\\
{\em IRAS}/25&	2.95e+01  $\pm$ 1.18e+00\\
{\em IRAS}/60	&4.07e+01 $\pm$  4.1e+00\\
{\em IRAS}/100&	1.83e+01  $\pm$ 2.0e+00\\
{\em AKARI}/9 &2.85e+00 $\pm$  1.68e-02\\
{\em AKARI}/18 &1.53e+01 $\pm$  1.20e-01\\
{\em AKARI}/65 &2.67e+01 $\pm$  2.50e+00\\
{\em AKARI}/90 &2.08e+01  $\pm$ 9.87e-01\\
{\em AKARI}/140 &6.88e+00  $\pm$ 9.03e-01\\
{\em WISE}/3.4 &1.44e-02  $\pm$ 3.05e-04\\
{\em WISE}/4.6& 1.13e-01 $\pm$  2.07e-03\\
{\em WISE}/12& 8.56e+00  $\pm$ 5.52e-02\\
{\em WISE}/22& 2.22e+01 $\pm$  1.02e-01\\
MIPS/24& 1.57e+01 $\pm$       8.01e-02\\
MIPS/70 &1.78e+01 $\pm$       1.85e-01\\
MIPS/160 & 2.82e+00 $\pm$       1.12e-01
\enddata
\tablenotetext{a}{$JHK$ photometry from \citet{2008A&A...479..817H}.}
\end{deluxetable}

\subsection{MIDI/VLTI}

Very Large Telescope Interferometric (VLTI) observations were obtained
of V605 Aql at the European Southern Observatory at Paranal in 2007
July using the MID-infrared Interferometric instrument (MIDI). The first night used the UT2--UT3 baseline, and
the second night used the UT4--UT3 baseline, but the source was apparently too large to generate fringes. {\bf This instrument is sensitive to structures with milliarcsecond sizes. Therefore, these data do not constrain the much larger structures detected elsewhere in this paper.} The MIDI spectrum
of V605\,Aql is shown in Figure~2. The photometric uncertainties are
about 20\%. The spectrum is featureless, and at a similar flux level
to a spectrum obtained with the Infrared Space Observatory (\/ISO),
more than 10 years previously.

\section{Direct Detection of the Central Star}

In Figure~3 (top panel), we show  the sum of the four polarimetric \HST\/ images, taken in
the intermediate-width F547M filter, which includes the strong stellar-wind
\ion{C}{4} emission feature. This is a very deep image, combining 6 \HST\/
orbits of imaging. Figure~3 is dominated by a
slightly non-stellar ``hot spot.''  The FWHM of this bright feature is
$0\farcs33$, as compared to $0\farcs18$ for nearby stars in the field. We
suggest that this is the central star itself, but still seen through substantial
extinction. The surrounding nebula is much fainter; in this filter the nebula is mostly starlight
scattered off dust plus a small amount of [\ion{O}{3}] 5007~\AA\
emission leaking in from the short-wavelength tail of the F547M bandpass.

\begin{figure}
\begin{center}
\includegraphics[width=2.0in,angle=-90]{f3a.eps}
\includegraphics[width=2.0in]{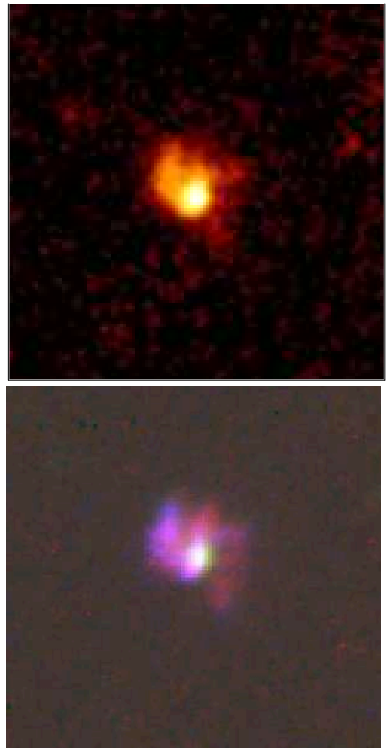}
\end{center}
 \figcaption{Images of V605~Aql.  {\em Top}: Deep \HST\/ image of
   in the F547M filter, obtained in 2009
   (WFPC2/WF2)\null. This bandpass is dominated by the \ion{C}{4}
   Wolf-Rayet emission feature in the spectrum of the central star.
    The image displays a bright, slightly extended ``hot spot,''
   which we interpret as the central star, seen through significant
   dust extinction.  {\em Bottom}: False-color rendition of the 2009
   \HST\/ images. Red is the nebular [\ion{N}{2}] emission
   line, green is the stellar \ion{C}{4} feature, and blue is the
   nebular [\ion{O}{3}] emission line. The putative central
   star, represented by the bright white spot, lies near the geometric
   center of the nebula. North is at the top, east on
   the left. Each panel is 5\arcsec x 5\arcsec.}
\end{figure}

To illustrate the spatial relationships between the different
features, we show in Figure~3 (bottom panel) a false-color image made by registering
and combining the [\ion{O}{3}], F547M, and [\ion{N}{2}] frames from
2009. This image demonstrates that our putative central star lies near
the geometric center of the compact nebula, including the faint
extension to the southwest.  Note that the near-IR {\it JHK\/} images
presented by \citet{2008A&A...479..817H} show the same nearly stellar
hot spot, which they argued is the central star.

\begin{figure*}
\begin{center}
\includegraphics[width=4in,angle=-90]{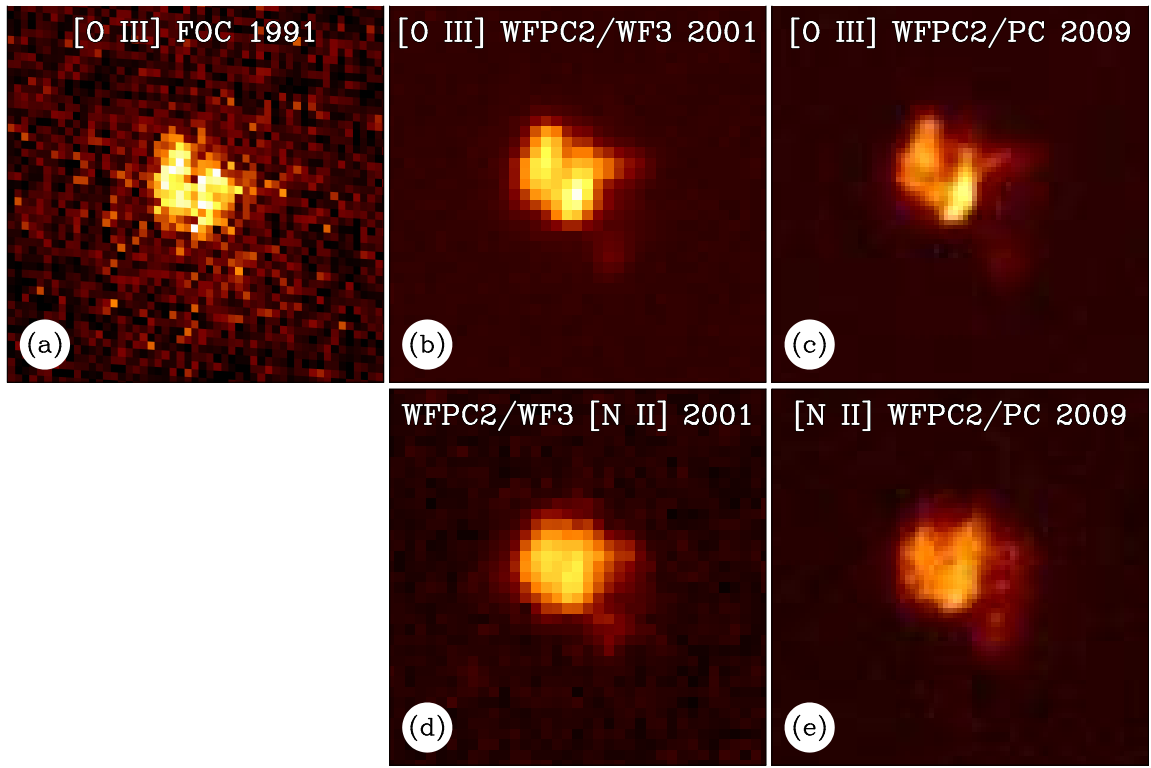}\\
\end{center} 
\figcaption{Top: \HST\/ images of the V605~Aql compact nebula in [\ion{O}{3}]
5007~\AA\ obtained in (a) 1991 (aberrated FOC), (b) 2001 (WFPC2/WF3), and (c) 2009
(WFPC2/PC)\null. Linear stretches were used. 
Bottom: \HST\/ images of V605~Aql in [\ion{N}{2}] 6583~\AA\ obtained in (d) 2001
(WFPC2/WF3) and (e) 2009 (WFPC2/PC)\null. 
Note expansion of the nebula from 2001 to 2009. Each panel is 5\arcsec x 5\arcsec.
North is at the top, east on the left in all images.}
\end{figure*}

Three of the four polarimetric F547M images were summed by combining
the various polarization angles into a Stokes {\it I} image using linear
combinations given by \citet{1997wfpc.rept...11B}. A ``standard" star
was chosen in the field which is listed in the GSC2.2 and UCAC4
catalogs as $V=15.57$, and in NOMAD at 15.4. We assumed a magnitude of
$V=15.5$ for the standard star. Photometry was done on this star and the
central star of V605 Aql in the F547M image, using standard techniques
in DAOPHOT and psf-fitting. We estimate that the V605 Aql central star
is $V=20.2\pm1.0$. The uncertainty is an estimate based on the
quality of the profile fitting and the uncertainty in the standard
star's magnitude, but does not try to estimate uncertainties
introduced by using a medium-pass filter with no color information.
If the the reddening estimate from \S5 is used, then the intrinsic
apparent brightness of V605 Aql in 2009 was $V=20.2-4\sim16.2$.
The F547M bandpass is dominated by the \ion{C}{4} Wolf-Rayet emission feature in the spectrum of the central star \citep{2006ApJ...646L..69C}. Therefore, we cannot use the photometry in this filter as an apparent stellar V magnitude which could be used with the estimated distance to calculate M$_V$.

\section{Morphology of the Compact Nebula}

Figure~4 shows the [\ion{O}{3}] images at the 1991, 2001, and 2009
epochs, and the [\ion{N}{2}] images from 2001 and 2009.  The brighter
portions of the compact nebula in 2009 have an angular diameter of
$\sim$$1''$; the total diameter is about 1\farcs7 if we include the
faint extensions on the southwest side.  \citet{2006ApJ...646L..69C}
suggested that V605 Aql may be surrounded by a thick torus of dust
with some stellar light scattering toward the observer from the poles
of the torus. This idea was discussed in detail by
\citet{2008A&A...479..817H}, based on their 2001 {\it HST\/} and
ground-based images.  This model was suggested by the earlier epochs
of imaging which showed two blobs of emission separated by a dark
band, thought to be an almost edge-on thick disk of dust obscuring the
central star in the visible.  \citet{1996ApJ...472..711G} first noted
the asymmetry of the central knot, and the dark band was already seen
in the aberrated FOC images from 1991 (Figure~4, top left panel), and
then more clearly in the deconvolved versions presented by
\citet{2002Ap&SS.279...31B}.  \citet{2008A&A...479..817H} suggested
that the central star was visible in their near-IR imaging and that
the disk was inclined enough to see the star in the center of the
disk. This morphology is suggestive of a bipolar structure possibly
with a central disk. The orientation of the dark band (PA $\sim$
150\degr) is also approximately parallel to the direction of the
major-axis of the surrounding PN, as shown in Figure~1.

Another interpretation of the morphology is that the northeast and
southwest emission is coming from material expanding away from the
star and its torus or disk at the center. This geometry is shown in
Figure~5. It is possible that the dust obscuration of the V605 Aql
central star and the nebula is decreasing with time so that in the
latest 2009 imaging, the central star is visible for the first time
and the southwest component is seen more clearly. The
northeast component is on the near side and less obscured than the
southwest component which is on the far side of the star. There is a gradient
in the ratio of [\ion{O}{3}] to [\ion{N}{2}] in the nebula, in the
sense that the ratio is highest on the northeast side and lowest on
the southwest. This could represent a range in dust extinction, lower
on the northeast, near-side, and higher on the southwest, far-side of
the star. The gradient could also be due to a range in excitation
level.

\begin{figure}
\centering
\includegraphics[width=2.25in,angle=0]{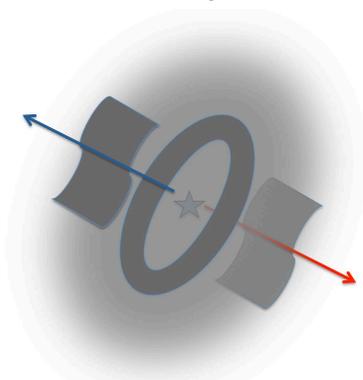}
\caption{Cartoon showing the central star of V605 Aql at the center of a tilted torus with the material in the northeast (upper left) expanding away on the near side, and the material in the southwest  (lower right) expanding away on the far side. The material on the far side experiences more dust extinction and so appears fainter.}
\end{figure}

\section{Emission-Line Spectral Modeling}

High-resolution (0.35~\AA) spectra of the integrated light of the compact nebula
were obtained in mid-1991 by \citet{1992MNRAS.257P..33P}, using the coud\'e echelle
spectrograph on the Anglo-Australian Telescope. These spectra showed a system of
broad and narrow emission-line components in [\ion{O}{3}] and [\ion{N}{2}], as
well as H$\alpha$ from the surrounding PN, Abell~58.   Unfortunately, the
original data have now been lost (D. Pollacco, private communication), so we
reconstructed them by digitizing the figures in the online version of the
journal article and converting them to tables of intensity vs.\ 
wavelength.

\citet{1992MNRAS.257P..33P} noted that H$\alpha$ and [\ion{N}{2}]
6548--6583~\AA\ have narrow emission-line components with double
profiles. For some unknown reason, the [O III] lines are not split.
Attributing these components to the surrounding PN,
they derived an expansion velocity of $31\pm4\,\kms$ and a systemic
velocity for the large PN of $+70\,\kms$.  The spectra also showed
much broader emission features in [\ion{N}{2}] and [\ion{O}{3}] (but
not in H$\alpha$), attributed to the compact hydrogen-deficient
nebula. The peaks of the broad features were blueshifted by 
$\sim$100 km~s$^{-1}$ with respect to the systemic velocity, and with FWHMs of
180 and 270 km~s$^{-1}$, respectively.  Pollacco et al.\ interpreted
these blueshifted features as arising either from a one-sided
collimated flow, or more probably from a dusty, approximately
spherical wind, in which the receding (i.e., redshifted) material is
obscured by dust within the wind.  The latter scenario is bolstered by
a subtle indication of a red tail in the broad components, and is the
one that we adopt here. 

We use a simple model of a spherical nebula of uniform density with
radius $R$, expanding at a constant speed $v_{\rm exp}$, and a
spherical $r,\theta$ coordinate system, where $\theta$ is the angle
relative to the line of sight. A volume element
located at $(r,\theta)$ will have radial velocity $v_r=-v_{\rm
  exp}\cos\theta$. If the optical depth to the center of the sphere is
$\tau$, the intensity from that volume element will be extinguished by
$\exp(-\tau[Z/R])$, where $Z=\sqrt{R^2-(r\sin\theta)^2}-r\cos\theta$
is the line-of-sight distance out of the nebula from the volume
element toward the observer.

In this model, the emission-line profile, $I(v_r)$, is an integral of
extinguished emission over the entire sphere.  An
asymmetric profile can thus be fit to the observed spectra using
standard Levenburg-Marquardt minimization \citep{1992nrfa.book.....P} of the
numeric integral as a function of systemic velocity, line width, and
the optical depth to the nebula center, $\tau$.  We performed such
fits on the [\ion{O}{3}] 5007\AA\ and [\ion{N}{2}]
6548--6583~\AA\ line complexes, as shown in Figure~6. The
doubled narrow components were forced to have no internal extinction,
as they represent emission from the outer PN, Abell~58.  The narrow lines yield a
mean systemic velocity of +80 km~s$^{-1}$, very close to that found by
\citet{1992MNRAS.257P..33P} from these spectra. The best fit model is also shown in Figure~6.

No single asymmetric profile can account for all of the broad
emission; however, we find that the bulk of the emission, including
the broad red and blue tails, can be well fit by a single sphere at a
systemic velocity of +80 km~s$^{-1}$, and an expansion velocity of 215
km~s$^{-1}$, with $\tau=4.0$ for [\ion{O}{3}] and 3.0 for
[\ion{N}{2}]. These components of the line profiles are shown in green
in Figure~6.  The ratio of optical depths is that
expected for pure carbon dust following a \citet[][hereafter MRN]{1977ApJ...217..425M} grain-size distribution, and corresponds to an
extinction toward the central star of $A_V=4$~mag.  Note that this
model does not account for profile broadening due to thermal
motions, slit width and the line-spread function of the optical
system.  To fully reproduce the observed profile, we would have to add
or subtract some narrow features, which suggests that the outflow has
clumpy substructure, consistent with the narrow-line images presented
in Figure~4.
The Galactic latitude of V605 Aql is $-5$\degr~which implies a
foreground interstellar extinction of $A_V\sim1.45 - 1.75$ mag
\citep{1998ApJ...500..525S,2011ApJ...737..103S}. 
So the total extinction toward V605 Aql is $\sim$5.5 mag. The uncertainty in this estimate is substantial as the extinction estimates from both the emission line fits and the foreground component contain simplifying assumptions.

\begin{figure}
\centering
\includegraphics[height=\linewidth,angle=-90]{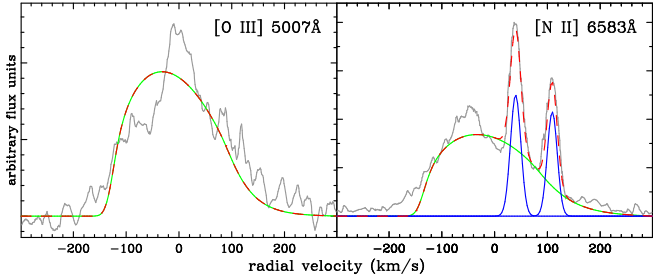}
\caption{ Model fits to the V605~Aql emission-line spectra of
  [\ion{O}{3}] 5007~\AA\ and [\ion{N}{2}] 6583~\AA\null. {\it Grey
    lines}: observed spectra from \citet{1992MNRAS.257P..33P}. {\it Blue
    profiles}:  [\ion{N}{2}] components attributed to the
  surrounding planetary nebula Abell~58, with center-of-mass velocity
  of $+75\,\kms$. {\it Green lines}: single-line profile fits using a
  uniform-density spherical nebula with a dust extinction of $A_V=4$
  to the center; these yield an expansion velocity $v_{\rm
    exp}=220\,\kms$ relative to center of mass.  {\it
    Dashed red lines}: sum of model components. 
 \label{linefit}}
\end{figure}

\section{Angular Expansion}

A primary goal of this investigation was to measure the angular expansion of the
compact nebula. Visual inspection of the
[\ion{O}{3}] and [\ion{N}{2}] images in Figure~4 does confirm noticeable
expansion. 

\begin{figure}
\centering
\includegraphics[height=\linewidth,angle=-90]{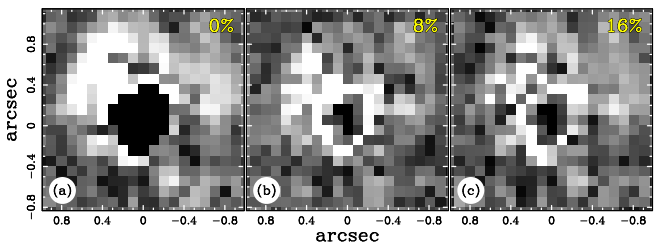}
\caption{
Difference images between the WFPC2 F502N data from 2001 and 2009, with the
former epoch magnified by the scaling shown in each panel. White is positive and black is negative. The images are centered on the star. (a)~Difference image
with no magnification, showing result of simple PSF-matching and flux-scaled
subtraction. (b)~Difference with 8\% magnification, yielding smallest sum of
residuals.  (c)~Difference with 16\% magnification.  Absolute residuals are
$\sim$5 times larger than for best-fit magnification of 8\%.
 \label{corrtile}}
\end{figure}

We chose the [\ion{O}{3}] images from 2001 and 2009 (separated by
7.8~years) for detailed analysis. As described in \S2, we resampled
the 2009 image to the (lower) resolution of the 2001 image. We then
magnified the earlier image by various factors and subtracted it from
the later one, and sought to minimize the residual fluxes.  Figure~7
shows PSF-matched difference images for magnifications of 0\%, 8\%,
and 16\% between the 2001 and 2009 epochs. The 0\% and 16\% difference
have much larger residuals due to the expansion of the nebula, and the
8\% image is the best fit.

The magnification residuals were minimized as follows. Assuming a homologous
expansion by a magnification factor $a$, and denoting the reference and input
images by $R$ and $I$, and the magnified input image by $aI$, we calculate the
sum of the squared  residuals, $\rho(a)=\sum_{\rm pix}(R-aI)^2$. The most likely
expansion factor is then the value $a_{\rm min}$ that minimizes $\rho$.

The uncertainty can be directly computed as $\sigma=\sqrt{\langle a^2 \rangle -
\langle a \rangle^2}$, where $\langle x \rangle$ is the expectation value of the
parameter $x$, i.e., $\langle a \rangle = \int{a\rho(a)da}$, if $\rho$ has been
normalized. When nebulae are better resolved, one can perform this task across
various wedges or quadrants in position angle, in order to determine the local
average expansion, or one can compare the expansion of individual features to
measure the microscopic expansion properties 
\citep{2002AJ....123.2676L,2006ApJ...641.1113U}.
However, since the V605~Aql nebula is only marginally resolved, we opted for a
global measurement in which the entire nebula is assumed to expand
homologously.  

As a note of caution, when the input and reference images have PSFs that differ
significantly, the residual $\rho$ may not be minimized for the actual best-fit
magnification factor, even if that factor is exactly known ahead of time.  We
attempted to minimize this problem by PSF-matching the WFPC2 images before
computing the difference residuals, and by masking out pixels surrounding the
bright, quasi-stellar source at the center of the nebula.

\begin{figure}
\centering
\includegraphics[height=\linewidth,angle=-90]{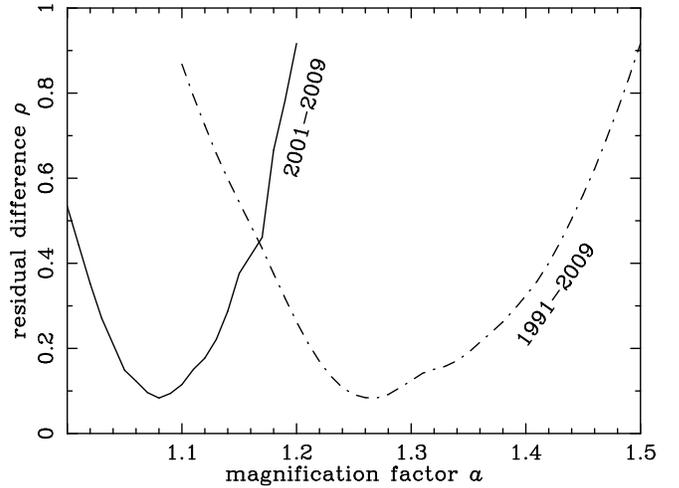}
\caption{Normalized residual differences $\rho(a)$ vs.\ magnification factor $a$
for the WFPC2 F501N images from 2001 and 2009 ({\it solid line}) and the
deconvolved 1991 F501N FOC and 2009 F502N WFPC2 images ({\it dot-dashed line}). 
The best-fit expansion of the nebula between the two epochs is the minimum of
each curve. 
 \label{corrfig}}
\end{figure}

Figure~8 plots the normalized residuals for the [\ion{O}{3}] filter
against magnification factor, for the 2001-2009 pair (solid line) and
the 1991-2009 pair (dot-dash line).  For the 2001--2009 WFPC2 images,
the derived expansion factor is $8\pm2\%$.  We verified this numerical
result in a ``double-blind'' fashion, by visually examining residual
images produced by scaling and subtracting the two epochs, and
selecting the magnification factors that appeared best, without
knowledge of the results plotted in Figure~8.  The flattest-appearing
residuals were indeed close to the 8\% value. Not shown here is a
similar exercise for the 2001--2009 [\ion{N}{2}] images, which gave an
expansion factor of $5\pm3\%$.

Under the assumption of homologous expansion, the 2001--2009
[\ion{O}{3}] data yield an ejection epoch of $1903\pm49$.  Performing
the same exercise between the the 2009 WFPC2 image and the FOC data
from 1991 yields an expansion factor of $27\pm 9\%$, as shown in
Figure~8, and an ejection epoch of $1926\pm 22$.  Both results are
thus consistent, within the errors, with an ejection of the compact
nebula at the time of the 1919 optical outburst.

\section{Geometric Distance}

We estimated the distance to V605\,Aql using the expansion-parallax method. In this technique, the distance is determined by equating the velocity implied by the angular expansion rate in the plane of the sky, derived in the previous two sections, to the expansion velocity in the line of sight. 

In order to estimate a distance to V605\,Aql based on the expansion parallax, we
need to convert the fractional expansion rate of the compact nebula into angular
units.  We measured this by considering each of the four quadrants of the image,
and determining the radial pixel shift needed to match features between epochs. 
We found that a radial shift of $0\farcs08\pm 0\farcs01$ minimizes the
subtraction residuals.

We also used a ``by-eye'' approach, in which we estimated the location of the
outer edge of the nebula from high-contrast renditions of the images. This
visual method also yielded an estimated expansion of $0\farcs08$ over the
7.8-year epoch difference.  The angular expansion rate is thus $10.2\pm1.3$
mas~yr$^{-1}$.  

A rough geometric distance, $d$, can be estimated from the observed angular
radius, $a\simeq0\farcs85$ at epoch $t=2009$, as discussed in \S4, combined
with the expansion velocity of $v_{\rm exp}=215\pm 20\,\kms$. We assume a
simple spherical expansion (in spite of the obviously non-spherical appearance
of the compact nebula), and that it was ejected in 1919. The result is

$$ d \simeq 4.8 \, {\rm kpc}
\left({0\farcs85}\over{a}\right)
\left({v_{\rm exp}\vphantom{d}}\over{215 \kms\vphantom{p}}\right)
\left({t-1919\vphantom{d}}\over{2009-1919\vphantom{p}}\right) \, ,
$$

\noindent but this is somewhat uncertain, since the nebular radius is poorly
defined.

An alternative calculation, using instead the angular expansion rate of 
$\dot{\theta}=10.2\pm 1.3\,\rm mas\,yr^{-1}$, yields

$$ d \simeq 4.4 \, {\rm kpc}
\left({v_{\rm exp}\vphantom{d}}\over{215 \kms\vphantom{p}}\right)
\left({10.2\,\rm mas \, yr^{-1}}\over{\dot\theta}\right) \, ,
$$

\noindent with a formal error of about $\pm$0.6~kpc. There is likely a
significant systematic error because of our assumption of a simple
spherical expansion. 
The images of the V605 Aql knot clearly show that it is not spherical so
the
expansion velocity in the plane of the sky may be larger than in the
radial direction, in which case our distance estimates would be too
low.

Other distance estimates, based on various methods applied to the large PN
Abell~58, have been reviewed by 
\citet{1997AJ....114.2679C} and \citet{2008A&A...479..817H}, and range from 2.7 to about 6~kpc. The result estimated here, $\sim$4.6~kpc, is consistent with the previous estimates, given the large uncertainties and systematic errors in all
of the methods.

\section{Dust Mass}

\citet{2001ApJ...559..419K} modeled the dust emission from V605 Aql
using a radiative transfer code which fit {\em Infrared Space
  Observatory} ({\it ISO}) and {\it IRAS} photometry of the star. They
assumed that $d = 5$ kpc, $T_{eff}$= 100,000 K, and $L_{\star}= 5800\
L_{\sun}$. They also assumed carbon dust with $A_V=7$ and found a
dust mass of $8\times10^{-3}\ M_{\sun}$.
\citet{2008A&A...479..817H} fit IR photometry from the J-band to 14.5
\micron~with two dust components emitting as blackbodies with
temperatures of 350 and 1500 K, but don't estimate the total dust
mass.

\citet{2008MNRAS.383.1639W} calculated a gas mass for the ejected knot
of $5\times10^{-5}\ M_{\sun}$ assuming a distance of 3.5 kpc,
$n_e= 2100$ cm$^{-3}$, and an angular radius of 0$\farcs$38. The
angular radius corresponds to $2\times 10^{16}$ cm at 3.5 kpc.
This agrees well with the size of a nebula expanding at 215 km
s$^{-1}$ for 90 yr ($6 \times 10^{16}$ cm).  They estimate the
extinction from c(H$\beta$) =2.0 for the inner knot which corresponds
to $A_V \sim 3$ mag.

The SED of V605 Aql, shown in Figure~2, was fit using various dust
components at different temperatures. As described in
\citet{2012ApJ...749..170S}, a zeroth-order estimate of the emitting
dust mass can be produced by fits to the observed SED.  Each dust
component has a single temperature, and emits as a blackbody modified
by the mass-absorption coefficient.  In addition, each was assumed to
be optically thin, 100\% amorphous carbon with an MRN distribution,
and optical constants from \citet{2001ApJ...548..296W}. Figure~2 shows
the best fit which required dust with three different temperatures and
masses (810 K , $1\times 10^{-11}\ M_{\sun}$; 235 K,
$9\times 10^{-6}\ M_{\sun}$; 75 K, $2\times 10^{-3}\ M_{\sun}$).
There is a range of models that fit the data roughly equivalently, with masses that are consistent to within a factor of two of those presented here.
Also, these dust masses are lower limits to the values one would
find when running a Monte-Carlo radiative-transfer model.

We have estimated in \S5 that V605 Aql suffers from $\sim$4 mag of internal
extinction at V. 
Let's assume a spherical
shell of dust with a radius of $2 \times 10^{16}$ cm around the star
as did \citet{2008MNRAS.383.1639W}. Then, looking through the dust to
the star, the optical depth is, $\tau=\kappa \rho L$, where $\rho$ is
the dust mass density, $\kappa$ is the mass absorption coefficient,
and L is the path length. V605 Aql is carbon rich
\citep{2006ApJ...646L..69C}, and the {\it Spitzer}/IRS spectrum of
shows no silicate or PAH features so the dust is likely to be
amorphous carbon \citep{2006MNRAS.373L..75E}.  A typical value for the
absorption coefficient is $\kappa \sim5 \times 10^{5}$ cm$^2$~g$^{-1}$
\citep{1996MNRAS.282.1321Z,2007ApJ...657..810D}. If $L = 2 \times
10^{16}$ cm and $\tau=4$, then $\rho = 4 \times 10^{-22}$
cm$^2$~g$^{-1}$.  Then, the mass of dust in the shell is just 4/3
$\pi L^3 \rho$ or $\sim 7 \times 10^{-6}\ M_{\sun}$.

This implies a gas-to-dust ratio of $\sim$7 which is very small compared to the typical gas-to-dust-ratio for the Galaxy of 100 \citep{2007ApJ...663..866D}. 
However, the gas-phase abundances in the central knot of V605 Aql are
very different than those seen in a typical interstellar
cloud. \citet{2008MNRAS.383.1639W} measured the fractional abundances
of various elements in the V605 knot, H (0.019), He (0.250), C
(0.021), N (0.043), O (0.323), and Ne (0.345). Firstly, H, which is
typically 75\% of interstellar gas, and does not deplete into dust
grains, is missing, which will lower the gas-to-dust ratio
significantly. Among the other abundant elements in the central knot,
Ne and He will also be undepleted as they do not participate in the
grain chemistry. In the ISM, C and O are heavily depleted into carbon
grains and silicate grains. However, the MIDI, {\em ISO}, and {\em
  Spitzer} Infrared Spectrograph (IRS) \citep{2006MNRAS.373L..75E}
observations of V605 Aql indicate that its mid-IR spectrum is a
featureless continuum. This is similar to what is seen in the RCB
stars \citep{2005ApJ...631L.147K}. There are no silicate features at
10 and 20 \micron, and no sign of PAHs or SiC. The featureless
spectrum indicates that the dust around V605 is probably amorphous
carbon, i.e., 100\% carbon. So, it is likely that carbon is highly
depleted into grains, but that the other elements, measured by
\citet{2008MNRAS.383.1639W}, are not. If carbon is depleted at a level
that we see in the ISM, then probably 90\% of the carbon is in dust.
In the gas phase in the V605 knot, carbon is $\sim$$1 \times 10^{-6}\
M_{\sun}$. If carbon is 90\% depleted, then the total mass of carbon
(gas phase and in dust) in the knot is $1 \times 10^{-5}\
M_{\sun}$. Then the gas-to-dust ratio is $\sim$5, which is remarkably
low. This estimate for the dust mass ($1 \times 10^{-5}\ M_{\sun}$)
fits in very well with the estimates above from the extinction and
likely size of the shell ($7 \times 10^{-6}\ M_{\sun}$), and from the
hot and warm dust components of the SED fit ($9 \times 10^{-6}\
M_{\sun}$). 

Assuming that the mass of C is actually $1 \times 10^{-5}\
M_{\sun}$, increases the mass of the central knot to $5.9 \times 10^{-5}\
M_{\sun}$, and changes the ratio C/O from 0.06 to 0.6 but still well below unity and far below the large C/O predicted for a FF. 

Wesson's estimate of the gas mass measures only hot gas near the star.
The second of the mass estimates in \citet{2008MNRAS.383.1639W} used a
density of 2100 cm$^{-3}$, derived from collisionally excited lines,
so that could underestimate the gas mass if there were regions of very
high density to quench the CEL emission.  Also, the apertures for the
near- and mid-IR photometry, plotted in Figure~2, have sizes of a few
arcseconds and are measuring the hot gas and dust in the central knot.
But, the apertures for the far-IR ($>$20 \micron) photometry are much
larger, {\em IRAS} (spatial resolution $\sim$4\arcmin) and {\em
  AKARI}/FIS (spatial resolution $\sim$35\arcsec), and may be
including colder dust further from the star, possibly associated with
the PN which is 44\arcsec~x 35\arcsec~in size
\citep{1966ApJ...144..259A}. This idea is supported by the fact that
the flux density in the {\it Spitzer}/MIPS images, which have a better
spatial resolution, is slightly lower.  So, the value estimated above,
$\sim 1 \times 10^{-5}\ M_{\sun}$, is probably consistent with the
dust mass in the central knot. The large mass of cold dust,
$2\times 10^{-3}\ M_{\sun}$, estimated from the SED fit, likely lies
outside the knot. A significant mass of cold dust was also recently found
around R CrB, which is a possible FF star
\citep{2011ApJ...743...44C}.

\section{Summary}

New polarimetric imaging with {\it HST\/}/WFPC2 in nebular [O III], [N II] and stellar C IV provides a baseline of 18 years to investigate the expansion of the ejecta from the 1919 event, and allow us to investigate the position and extinction of the central star. The new measurements of the rate of the expansion, reported here, are consistent with an ejection in 1919. When combined with velocities measured from  spectra of V605 Aql, a distance of 4.6 kpc is estimated, which is in the middle of the range of previous distance estimates. 

The morphology of the 1919 ejecta is highly non-spherical. It seems to be either a tilted dusty torus with a dark band, or a disk surrounding the central star with bipolar ejections on either side. Further imaging as the expansion continues may clarify this. The asymmetry of the knot is aligned with the asymmetry of the much larger surrounding PN. A disk and PN asymmetry could result if V605 Aql is a binary system \citep{2011MNRAS.410.1870L}. There is early evidence that there is a disk around Sakurai's Object \citep{2009A&A...493L..17C}. 

The new {\it HST\/} images of V605 Aql, presented here, provide the
first direct detection of the central star in the visible since
1923. In the stacked F547M image, the star is $V= 20.2 \pm 1.0$
mag. We estimate that the intrinsic apparent magnitude today should be
$V \sim 16.2$ mag, assuming the extinction is A$_V$$\sim$ 4 mag which
is estimated from the spectral modeling. This modeling estimates a
systemic velocity of +80 km s$^{-1}$ and an expansion velocity of 215
km s$^{-1}$.  Polarimetric imaging was obtained to investigate whether
the 2001 spectrum of V605 Aql \citep{2006ApJ...646L..69C} was obtained
primarily in scattered light from dust in the central knot, but the
signal-to-noise in the data was not sufficient to measure the level of
polarization.

It is found that the SED of the dust around V605 Aql, consisting of
Near-IR, IRAS, WISE, AKARI, and Spitzer photometry, can be fit using
carbon dust with with three different temperatures and masses. Using
three different methods, we estimate a mass of warm dust in the
central knot of $\sim$10$^{-5}\ M_{\sun}$. A significant mass of cold
dust ($10^{-3}\  M_{\sun}$) may be spread over a larger volume perhaps
associated with the surrounding PN.
 
V605 Aql is one of only three stars, identified with a final helium-shell flash, to be observed from their initial brightening. It is very similar to the more recent FF star, Sakurai's Object. Monitoring of V605 Aql is important because it may allow the future behavior of Sakurai's Object to be predicted and better observed. 
 In V605 Aql, the composition of the hydrogen-deficient ejecta (the C/O ratio below unity and a Ne mass fraction of 35\%), however, leaves little doubt that the scenario that formed this object is far more complex than a FF (that predicts C/O far larger than unity and a Ne mass fraction of ~2\% \citep{2006PASP..118..183W}). V605 Aql  must have passed through a phase involving the dredge-up of Ne from the core of an ONe WD. Another object currently listed in the FF class, Abell 30, also has hydrogen-deficient knots with similar abundance trends (C/O below unity and a Ne mass fraction between 8 and 20\% \citep{2003MNRAS.340..253W}). The ONe nova scenario, proposed by \citet{2011MNRAS.410.1870L} to explain these abundance patterns, may not be viable because it does not include the compulsory series of carbon-oxygen nova outbursts that precede the ONe nova detonation. If so, then the only viable scenario is that of a merger between a massive ONe white dwarf and a companion. At the present time, it is not possible to resolve whether V605 Aql is the result of a final helium-shell flash, an ONe nova, or a merger event.

\acknowledgments

We thank the anonymous referee for several helpful suggestions. Support for this work was provided by NASA through grant number HST-GO-11985 from
the Space Telescope Science Institute, which is operated by AURA, Inc., under
NASA contract NAS 5-26555; and by the Goucher College Office of the Provost. 
Some of the data presented in this paper were obtained from the Mikulski Archive
for Space Telescopes (MAST). We would like to acknowledge James Liebert, Alvio Renzini, and Nye Evans for assisting with this work. This research has made use of the VizieR catalogue access tool, CDS, Strasbourg, France.

{\it Facilities:} \facility{HST (WFPC2, FOS, FOC)}

\bibliography{/Users/gclayton/projects/latexstuff/everything2}


\end{document}